# From Failure Modes to Reliability Awareness in Generative and Agentic AI System

Janet (Jing) Lin[1], Liangwei Zhang[2]

[1]Division of Operation and Maintenance, Luleå University of Technology, Luleå, Sweden
[2]Department of Industrial Engineering, Dongguan University of Technology, Dongguan, China

**Abstract** This chapter bridges technical analysis and organizational preparedness by tracing the path from layered failure modes to reliability awareness in generative and agentic AI systems. We first introduce an 11-layer failure stack, a structured framework for identifying vulnerabilities ranging from hardware and power foundations to adaptive learning and agentic reasoning. Building on this, the chapter demonstrates how failures rarely occur in isolation but propagate across layers, creating cascading effects with systemic consequences. To complement this diagnostic lens, we develop the concept of awareness mapping: a maturity-oriented framework that quantifies how well individuals and organizations recognize reliability risks across the AI stack. Awareness is treated not only as a diagnostic score but also as a strategic input for AI governance, guiding improvement and resilience planning. By linking layered failures to awareness levels and further integrating this into Dependability-Centred Asset Management (DCAM), the chapter positions awareness mapping as both a measurement tool and a roadmap for trustworthy and sustainable AI deployment across mission-critical domains.

## 1 Introduction

AI systems are increasingly deployed in safety- and mission-critical domains such as transportation, energy, healthcare, manufacturing, and the built environment. Reliability has long been a central concern for what we call conventional AI systems — machine learning applications designed for tasks such as classification, prediction, optimization, and control. These systems face multi-level vulnerabilities, spanning hardware stability, power quality, data integrity, model robustness, and application integration.

The arrival of generative AI expands this landscape. By producing open-ended outputs — text, images, code, or designs — generative systems introduce new reliability challenges, including hallucinations, factual errors, toxic or biased content, and potential misuse through deepfakes or disinformation. These risks extend beyond conventional performance concerns to questions of trustworthiness, safety, and governance (Joshi, 2025).

The next wave, agentic AI, deepens the challenge further. Combining autonomy, planning, reasoning, memory, and multi-agent interaction, agentic systems can pursue goals and initiating actions with system-wide consequences. Their failure modes include goal misalignment, flawed planning, emergent conflicts between agents, and breakdowns in human–AI collaboration (Acharya, Kuppan and Divya, 2025).

This chapter addresses these paradigms in continuity, asking: *where and how can conventional, generative, and agentic AI systems fail — and how aware are we of these vulnerabilities?* To answer, the chapter introduces two complementary contributions:

- The 11-layer failure stack, a structured framework tracing vulnerabilities from physical computation and energy through data, models, and applications, up to learning, reasoning, and multi-agent coordination.
- The concept of awareness mapping, which assesses how well individuals and organizations recognize risks across these layers and positions awareness itself as a dimension of reliability.

Case vignettes drawn from transportation, energy, healthcare, manufacturing, and the built environment illustrate both layer-specific vulnerabilities and cascading, cross-layer effects. Finally, the chapter situates these tools within the paradigm of Dependability-Centred Asset Management (DCAM), linking technical failure analysis to lifecycle strategies for trustworthy and sustainable deployment of generative and agentic AI systems.

## 2 Reliability as a Moving Target in Conventional, Generative, and Agentic AI

Reliability has traditionally been treated as something that could be designed for and verified — a destination rather than a journey. In practice, however, even in classical engineering domains, reliability has always been a moving target. It shifts over time with operating conditions, usage patterns, maintenance practices, and unexpected interactions. What was once "reliable" in a test environment may not hold under long-term use or in new contexts.

This dynamism becomes even more pronounced in AI systems. Unlike physical components whose degradation can often be modelled in predictable ways, AI systems continuously interact with changing data, evolving environments, and — in the case of agentic AI — with other agents and human stakeholders. Reliability here is not a static property but a dynamic relationship, shaped by macro-level regulations and institutions, meso-level organizational practices, and micro-level component behaviours (Lin and Silfvenius, 2025).

To ground this discussion, we turn first to established definitions. International standards such as ISO, IEC, and IEEE define reliability as: "*The ability of a system or component to perform its required functions under stated conditions for a specified period of time.*" (Zhang *et al.*, 2017)

This classical definition provides a solid foundation, but its meaning shifts as we move from traditional physical and software systems to conventional AI, generative AI, and agentic AI. Each paradigm forces us to reinterpret what counts as the "intended function" and which vulnerabilities matter most.

---

**Box 1. Reliability: From Standards to AI Paradigms**

**Standard definition (ISO/IEC, IEEE, MIL-STD):**

*"Reliability is the ability of a system or component to perform its required functions under stated conditions for a specified period of time."*

- **Traditional systems:** Intended functions are predictable and bounded, failures arise from wear-out, material degradation, or software bugs.
- **Conventional AI systems:** Intended functions involve consistent and accurate predictions, classifications, or optimizations, dependent on data pipelines and computational infrastructure.
- **Generative AI systems:** Intended functions expand to producing trustworthy, safe, and contextually appropriate outputs such as text, images, or code.
- **Agentic AI systems:** Intended functions extend further to reasoning, planning, adapting, and safely interacting with humans and other agents in open-ended environments.

**Unified perspective:**

*Reliability is the dependable functioning of a system across its full stack — from physical infrastructure and data pipelines to models, outputs, reasoning, and interactions — ensuring that intended functions are performed accurately, safely, and sustainably over time under evolving conditions.*

---

This progression demonstrates that reliability is cumulative. Each paradigm inherits the concerns of the ones below — physical durability, computational stability, data and model robustness — while introducing new dimensions shaped by its scope. Conventional AI depends on reliable infrastructure but adds sensitivity to data and models. Generative AI builds on these layers while demanding content reliability and safety. Agentic AI inherits all the above, extending reliability into goal alignment, adaptive learning, and emergent multi-agent behaviour (Fig. 1).

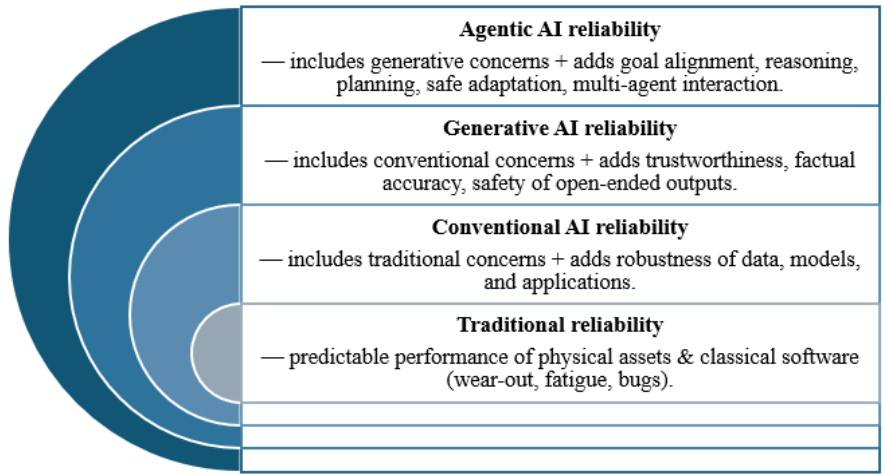

**Fig. 1.** Reliability as an expanding, cumulative concept. Each paradigm inherits the reliability concerns of the ones below while introducing new dimensions shaped by its scope. Traditional reliability emphasizes physical and software dependability; conventional AI adds data and model robustness; generative AI extends to trustworthy content; and agentic AI expands further into reasoning, adaptation, and interaction.

In this sense, AI does not change the fact that reliability is a moving target — it amplifies it. By redefining what counts as an "intended function," AI systems shift the trajectory of reliability challenges and require continuous adaptation in engineering practice (Table 1). At the same time, the very features that complicate reliability — adaptivity, perception, and reasoning — also create new opportunities, from advanced diagnostics to self-healing mechanisms. AI makes reliability both harder and more achievable: harder, because it multiplies potential failure modes; more achievable, because it equips us with intelligent tools to anticipate and mitigate them in real time.

**Table 1** Evolving interpretations of reliability across system paradigms

| System Type | How the Standard Definition Applies | Characteristic Failure Emphases |
|---|---|---|
| Traditional systems (Physical assets & classical software) | *Performing intended functions under stated conditions* = predictable operation of components or code over time, managed by design quality, testing, and maintenance. | Wear-out, fatigue, corrosion, software bugs, maintenance errors. |
| Conventional AI systems | *Intended function* = providing consistent and accurate predictions, classifications, or optimizations, supported by reliable infrastructure. | Data drift, mislabelling, model instability, hardware/software faults, silent monitoring failures. |
| Generative AI systems | *Intended function* = producing trustworthy, factual, safe, and contextually appropriate outputs, stable at scale. | Hallucinations, factual inconsistency, bias, unsafe content, misuse (deepfakes), scaling failures. |
| Agentic AI systems | *Intended function* = reasoning, planning, adapting, and safely pursuing goals in dynamic, multi-agent contexts. | Goal misalignment, flawed reasoning, unsafe adaptation, emergent conflicts, cascading systemic failures. |

Yet AI is not only an amplifier of existing reliability dynamics. It can also shift or misdirect the trajectory of how reliability is understood and managed. By redefining what counts as reliable

behaviour — for example, emphasizing output plausibility rather than operational stability — AI risks encouraging organizations to underemphasize traditional dependability concerns or overestimate AI's self-correcting capacities. As argued in earlier work on the intrinsic mechanisms of reliability improvement (Lin and Silfvenius, 2025), reliability must be seen as a continuously evolving system property, requiring deliberate and ongoing enhancement. This perspective reinforces why a structured, layered approach is essential: it not only catalogs failure modes but also provides guidance for sustaining and improving reliability in the era of conventional, generative, and agentic AI.

Thus, the standards-based definition remains essential, but for AI systems it must be extended into a layered and dynamic conception of reliability. This prepares the ground for Section 3, where we introduce the 11-layer failure stack and map how vulnerabilities manifest differently across conventional, generative, and agentic AI systems.

## 3. Layered Failure Modes Across Conventional, Generative, and Agentic AI Systems

Failures in AI systems rarely occur in isolation. They emerge from multiple interacting layers, spanning from physical computation and infrastructure through data pipelines, models, and applications, up to adaptive learning and multi-agent coordination. To capture this complexity, we adopt an 11-layer failure stack: a structured framework that traces potential vulnerabilities across the full spectrum of AI systems.

This failure stack provides a baseline lens for analysing reliability challenges. Yet its manifestations are not uniform. In conventional AI, failures often concentrate around data quality, model robustness, and integration into domain applications. In generative AI, reliability hinges on the factuality, appropriateness, and safe use of generated content. In agentic AI, challenges extend further into goal alignment, reasoning fidelity, adaptation in dynamic environments, and emergent multi-agent behaviours.

To organize this analysis, the section unfolds in three steps:
- 3.1 introduces the 11-layer failure stack as a general framework, providing an overview of its structure and logic.
- 3.2 develops a comparative view, showing how each layer manifests differently across conventional, generative, and agentic AI systems.
- 3.3 offers paradigm spotlights, highlighting distinctive reliability challenges and illustrative vignettes that demonstrate how failures unfold in real-world contexts.
- 3.4 draws the threads together by examining cross-layer risks and cascading effects, showing how localized faults propagate through the stack and amplify into systemic failures.

Taken together, this layered perspective reveals both shared vulnerabilities and paradigm-specific risks, underscoring that AI reliability cannot be safeguarded at a single layer alone. It requires a cross-layer, paradigm-sensitive approach that adapts to the evolving nature of AI systems (Ale *et al.*, 2025).

### *3.1 The 11-Layer Failure Stack: A General Framework*

Ensuring the reliability of AI systems requires more than verifying isolated components. Failures often arise from complex interactions across multiple layers, where vulnerabilities at one level propagate upward or downward, creating systemic risks. To capture this complexity, we propose an 11-layer failure stack, which organizes potential vulnerabilities from the physical substrate of computation to the highest levels of reasoning, adaptation, and multi-agent interaction.

The 11 layers can be grouped into four broad domains:
1. Foundational layers (hardware, power, system software, AI frameworks) — the physical and computational substrate.
2. Core intelligence layers (models and data) — the learning and knowledge base.

3. Operational layers (applications, execution, monitoring, learning) — the deployment and adaptation mechanisms.
4. Agentic layer (reasoning, goal alignment, multi-agent coordination) — the level where autonomy and decision-making unfold.

Each layer has a distinct role, typical failure modes, and sectoral manifestations. Table 2 provides a high-level summary.

**Table 2** Overview of the 11-Layer Failure Stack

| Layer | Function | Representative Failure Modes |
| --- | --- | --- |
| **1. Hardware** | Computing substrate: processors, memory, storage, interconnects | Overheating, memory bit flips, device wear-out, interconnect degradation |
| **2. Power & Energy** | Stable power supply: PSUs, UPS, energy management | Voltage instability, surges/spikes, thermal overload, battery depletion |
| **3. System Software** | OS, drivers, virtualization, firmware | Kernel panics, driver incompatibility, firmware bugs, virtualization overhead |
| **4. AI Frameworks** | ML/DL libraries and pipelines | Dependency conflicts, numerical instability, non-deterministic training |
| **5. Models** | Encoded AI knowledge and decision rules | Overfitting, underfitting, adversarial attacks, hallucinations |
| **6. Data** | Input pipelines, storage, labelling | Data drift, noise, mislabelled data, integration errors, unreliable synthetic data |
| **7. Applications** | Domain-specific AI use cases | API downtime, poor integration, UI errors |
| **8. Execution** | Real-time orchestration: cloud/edge, load balancing | Latency spikes, cold starts, resource starvation |
| **9. Monitoring** | Observability, logging, alerts | Silent failure, undetected drift, alert fatigue |
| **10. Learning** | Continuous learning and adaptation | Concept drift, misaligned feedback loops, retraining on biased data |
| **11. Agentic AI** | Reasoning, planning, multi-agent coordination | Goal misalignment, emergent conflicts, opaque decision-making |

**Why Layers Matter Together**

The 11-layer stack is not just a checklist—it emphasizes that failures are layered and interdependent. A data error (Layer 6) can cascade into faulty models (Layer 5), misguide applications (Layer 7), and distort agentic coordination (Layer 11). Conversely, poor reasoning at the agentic level can impose stress on infrastructure, amplifying faults downward. Thus, the 11-layer stack provides both:

- a diagnostic lens (where can failures occur?), and
- a design principle (how should systems be architected to contain cascading risks?).

In the next section (3.2), we extend this framework into a comparative analysis, showing how the same layers manifest differently across conventional, generative, and agentic AI systems.

### 3.1.1 Layer 1: Hardware

The hardware layer provides the physical foundation for all AI systems. It encompasses processors (CPUs, GPUs, TPUs), memory, storage devices, and interconnects that carry out and sustain computation. Without reliable hardware, higher levels of the AI stack cannot operate correctly, as every model, data flow, and decision ultimately depend on the integrity of the physical substrate.

Hardware failures often emerge from:

- Thermal stress, such as overheating in processors or accelerators.
- Wear-out and fatigue, including degraded solder joints or storage media failure.
- Transient faults, such as memory bit flips from cosmic radiation or electromagnetic interference.
- Environmental stress, including vibration, dust, or humidity damaging sensitive circuits.

These failures can be either catastrophic (system crashes) or silent (bit errors that propagate unnoticed).

> **Box 2. Cross-Sector Examples of Layer 1: Hardware**
>
> - **Transportation (railways, aviation):** onboard AI accelerators in trains overheat during summer heatwaves, causing intermittent diagnostic failures; avionics hardware experiences accelerated aging under pressure and vibration.
> - **Energy (power grid):** edge devices in substations are corrupted by voltage spikes, leading to AI controllers missing grid instability signals.
> - **Healthcare:** GPUs used in MRI imaging exhibit memory corruption, silently distorting image analyses and leading to diagnostic errors.
> - **Manufacturing:** robot controller boards fail due to vibration and dust on the shop floor, halting production lines unexpectedly.
> - **Buildings (HVAC):** controllers for adaptive HVAC systems fatigue under continuous switching cycles, degrading comfort and efficiency.

Classical hardware reliability engineering has a long tradition of mitigation strategies, including:
- Redundancy (backup processors, fault-tolerant architectures).
- Error-Correcting Codes (ECC) to detect and correct memory errors.
- Thermal management through active cooling and environmental design.

However, the demands of AI hardware introduce new challenges. Accelerators like GPUs and TPUs operate at much higher utilization and parallelism, making them more susceptible to thermal cycling and component stress. Neuromorphic and edge-AI chips present emerging failure behaviours that are less well understood. Research is exploring self-monitoring hardware, adaptive voltage/frequency scaling, and AI-driven prognostics to predict hardware degradation (Shankar and Muralidhar, 2025). Yet, deployment in safety-critical contexts such as healthcare or transportation remains limited.

**Link to Higher Layers**

Hardware failures often propagate invisibly. A silent memory error at this level can corrupt model weights (Layer 5), leading to systematic misclassifications. A processor glitch can trigger operating system crashes (Layer 3), cascading into downtime at the application or agentic layers. Thus, ensuring robustness at the hardware layer is not only foundational but also essential for preserving reliability throughout the entire AI stack.

### *3.1.2 Layer 2: Power & Energy*
The power and energy layer supplies the stable electrical foundation that enables AI hardware to function. This includes power supply units (PSUs), batteries, uninterruptible power supplies (UPS), and energy management systems. Reliability here ensures not only continuous operation but also protection against fluctuations and interruptions that can compromise computation (Huang *et al.*, 2025).

Power & Energy failures often emerge from:
- Voltage instability due to grid fluctuations.
- Thermal overload of power components.
- Surge or spike damage from transient events.
- Battery depletion in mobile or edge devices.
- Grounding or wiring faults in deployed environments.

> **Box 3. Cross-Sector Examples of Layer 2: Power & Energy**
>
> - **Transportation:** autonomous inspection robots in railways cut short missions due to rapid battery drain; EV onboard AI controllers reset when encountering voltage dips.
> - **Energy:** smart grid AI controllers malfunction during frequency fluctuations, delaying corrective actions.
> - **Healthcare:** portable diagnostic AI devices shut down mid-operation due to unstable supply or battery failure.
> - **Manufacturing:** AI-driven CNC machines halt when UPS systems fail during local power dips.
> - **Buildings (HVAC):** controllers reset during building-wide power transitions, disrupting environmental stability.

Reliability engineering for power systems emphasizes redundant power supplies, surge protection, and battery management systems. In data centers, advanced energy-aware scheduling balances AI workload against power draw. In edge-AI, low-power chips and adaptive energy management extend device life. Yet, as AI systems move into mission-critical contexts, even minor instabilities are unacceptable. Research is shifting toward AI-enabled power monitoring (predicting battery life, detecting instability) and integration of renewable energy sources, though these approaches are not yet mature for safety-critical deployment.

**Link to Higher Layers**

Failures here propagate rapidly upward. A transient power loss can cause firmware corruption (Layer 3), data pipeline interruption (Layer 6), or unexpected resets that undermine multi-agent coordination (Layer 11). Reliable AI requires stable power as a non-negotiable foundation.

### *3.1.3 Layer 3: System Software*

System software sits between hardware and higher-level frameworks, providing the operating environment for AI workloads. It includes operating systems, device drivers, virtualization layers, and firmware. This layer ensures that physical components are accessible and stable for applications above.

Typical System Software Failures include (Ebad, 2018):

- Kernel panics or OS crashes under load.
- Driver incompatibility with accelerators (e.g., GPU drivers).
- Firmware bugs that destabilize control systems.
- Virtualization overhead or clock skew, affecting real-time tasks.
- Poorly timed updates introducing regressions.

> **Box 4. Cross-Sector Examples of Layer 3: System Software**
>
> - **Transportation:** a driver updates causes SCADA-based railway monitoring to crash, delaying predictive maintenance.
> - **Energy:** a firmware bug in inverter control loops destabilizes renewable grid integration.
> - **Healthcare:** surgical robots stall when GPU driver crashes mid-operation.
> - **Manufacturing:** a PLC firmware bug halts robotic arm coordination.
> - **Buildings (HVAC):** building automation freezes after an OS update conflict.

System software reliability is usually addressed through certification, long-term support kernels, and redundant control firmware. Virtualization adds isolation but also complexity, raising new risks. In AI, the challenge is heightened because drivers and firmware must keep pace with rapidly evolving hardware accelerators. Research explores lightweight hypervisors, formally verified kernels, and continuous integration pipelines for system software. However, many industries still face gaps between rapidly updated AI stacks and conservative operational environments (e.g., energy or healthcare).

**Link to Higher Layers**

System software is the glue between physical hardware and AI frameworks. Failures here can ripple upward, rendering frameworks unusable (Layer 4), halting model inference (Layer 5), or causing application downtime (Layer 7). Ensuring resilience at this layer requires balancing stability and adaptability.

### *3.1.4 Layer 4: AI Frameworks*

AI frameworks provide the libraries and pipelines that enable developers to build, train, and deploy models. This includes deep learning platforms (e.g., TensorFlow, PyTorch), optimization toolkits, and inference runtimes. Frameworks standardize access to hardware accelerators and simplify large-scale model training, but they also introduce dependencies and complexity (Weber, 2022).

Typical AI Frameworks Failures include:
- Dependency conflicts between framework versions.
- Numerical instabilities during training (e.g., exploding/vanishing gradients).
- Non-deterministic behaviour, making models difficult to reproduce.
- Poor backward compatibility when upgrading frameworks.

> **Box 5. Cross-Sector Examples of Layer 4: AI Frameworks**
> 
> - **Transportation:** a mismatch in TensorFlow versions breaks predictive maintenance deployment in railway monitoring systems.
> - **Energy:** instability in an optimization tool degrades renewable energy forecasting accuracy.
> - **Healthcare:** PyTorch inference errors disrupt medical image segmentation workflows.
> - **Manufacturing:** predictive maintenance pipelines fail after a framework upgrade, halting rollout.
> - **Buildings (HVAC):** deep learning classifiers for fault detection cannot run due to unresolved framework bugs.

Current best practices emphasize containerization, dependency pinning, and continuous integration testing. Open-source ecosystems like PyTorch and TensorFlow evolve rapidly, enabling cutting-edge applications but creating instability for safety-critical industries. Research focuses on deterministic training, lightweight inference runtimes, and framework certification for high-assurance domains. Adoption in regulated industries remains slow, as frameworks are rarely validated for dependability.

**Link to Higher Layers**

If frameworks fail, models (Layer 5) cannot run or reproduce reliably, undermining application-level dependability (Layer 7). This layer thus acts as a keystone connecting computational resources to usable intelligence.

### *3.1.5 Layer 5: Models*

Models encode the learned intelligence of AI systems, transforming data into predictions, classifications, or decisions. This includes conventional ML models, deep neural networks, foundation models, and large language models (LLMs).

Typical Models Failures include:
- Overfitting or underfitting, reducing generalizability.
- Adversarial vulnerabilities, where small perturbations trigger misclassification.
- Hallucinations, particularly in generative models producing plausible but false outputs.
- Model drift, as performance degrades under new data distributions.

> **Box 6. Cross-Sector Examples of Layer 5: Models**
>
> - **Transportation:** bogie fault classifiers misdiagnose rare weather-related vibrations.
> - **Energy:** renewable generation predictors fail during extreme storms.
> - **Healthcare:** LLM-based assistants hallucinate treatment recommendations.
> - **Manufacturing:** defect detection models miss rare but critical cracks in products.
> - **Buildings (HVAC):** occupancy predictors misclassify patterns, wasting energy.

Research on robustness, fairness, and explainability has grown rapidly. Techniques such as adversarial training, calibration methods, and uncertainty quantification aim to improve reliability. Yet, the rise of foundation and generative models introduces new risks: hallucinations, misaligned goals, and opaque decision-making (T. Zhang *et al.*, 2025). Current progress includes fine-tuning guardrails, alignment methods such as Reinforcement Learning from Human Feedback (RLHF), and certification efforts, but practical assurance for mission-critical use is limited.

**Link to Higher Layers**

Model failures propagate directly to the application layer (7) and can misguide multi-agent coordination (Layer 11). Because models form the intelligence core, reliability at this layer is highly visible to end-users, often overshadowing vulnerabilities in lower layers.

### *3.1.6 Layer 6: Data*

Data forms the lifeblood of AI systems. This layer includes data pipelines, storage, collection devices, labelling infrastructures, and synthetic data generation. Reliable AI depends on the quality, integrity, and timeliness of data.

Typical Data Failures include:
- Data drift, where input distributions shift over time.
- Noise or corruption from faulty sensors.
- Labelling errors, especially in supervised learning.
- Integration errors when fusing multiple data sources.
- Synthetic data unreliability, if generation processes introduce hidden biases.

> **Box 7. Cross-Sector Examples of Layer 6: Data**
>
> - **Transportation:** misaligned vibration sensors cause false rail track degradation alerts.
> - **Energy:** poor synchronization between SCADA and PMU data corrupts forecasting.
> - **Healthcare:** mislabelled medical images bias diagnostic models.
> - **Manufacturing:** inconsistent calibration across assembly lines leads to unreliable datasets.
> - **Buildings (HVAC):** faulty $CO_2$ sensors skew ventilation control algorithms.

Data reliability research emphasizes data validation pipelines, anomaly detection, and active data curation (Sharma, Kumar and Kaswan, 2021). Advances in synthetic data generation promise coverage of rare events but risk introducing unrepresentative distributions. Industry practice often underestimates the difficulty of maintaining high-quality, real-time data pipelines, especially when integrating legacy sensors. Ongoing research explores data provenance tracking and AI-driven labelling quality assurance.
**Link to Higher Layers**
Faulty data undermines model training (Layer 5), disrupts applications (Layer 7), and erodes monitoring reliability (Layer 9). In many cases, data-layer failures propagate invisibly, making this one of the most critical and underestimated points of failure in AI systems.

### *3.1.7 Layer 7: Application*
The application layer represents the domain-specific integration of AI into workflows and real-world decision-making. It connects models to operators, assets, and end-users through dashboards, APIs, or automation systems. This is where AI's predictions become actionable.
Typical Application Failures include:
- Poor integration with legacy systems.
- API downtime that interrupts functionality.
- User interface errors leading to misinterpretation of outputs.
- Context mismatch, where AI decisions are applied inappropriately.

> **Box 8. Cross-Sector Examples of Layer 7: Application**
>
> - **Transportation:** predictive maintenance scheduling fails when APIs go offline, disrupting fleet operations.
> - **Energy:** load-balancing tools misallocate resources during peak demand due to poor integration.
> - **Healthcare:** clinical decision support tools crash mid-workflow, delaying treatment.
> - **Manufacturing:** AI scheduling software misallocates machine resources, causing production bottlenecks.
> - **Buildings (**HVAC): dashboards misreport ventilation status, leading to operator misjudgements.

Application reliability is supported through API testing, modular integration, and resilience design (Kothamali, 2025). However, in practice, applications often depend on brittle middleware and poorly monitored interfaces. Current research emphasizes human-centred design, trust-calibrated interfaces, and explainable outputs to strengthen the reliability of AI-assisted decision-making**.**
**Link to Higher Layers**
Application failures often obscure root causes: operators may blame the application even when the true fault lies in data (Layer 6) or models (Layer 5). This makes robust application design critical as the final touchpoint between AI and human trust.

### *3.1.8 Layer 8: Execution*
The execution layer governs the real-time orchestration of AI workloads. This includes cloud/edge scheduling, parallel execution, and load balancing (Alsadie and Alsulami, 2024). Reliability here ensures that AI models and applications can run under constraints of latency, scale, and computational resources.
Typical Failures include:
- Latency spikes degrading real-time performance.
- Cold starts in serverless architectures causing delays.
- Resource starvation when workloads exceed system capacity.
- Communication bottlenecks between distributed nodes.

> **Box 9. Cross-Sector Examples of Layer 8: Execution**
>
> - **Transportation:** railway dispatch AI lags, propagating train delays.
> - **Energy:** cloud inference delays undermine fast frequency response in smart grids.
> - **Healthcare:** triage systems experience inference latency, delaying emergency alerts.
> - **Manufacturing:** robot coordination lags, stopping assembly lines.
> - **Buildings (HVAC):** edge controllers' overload when rnning multiple fault checks simultaneously.

Solutions include container orchestration (Kubernetes), real-time schedulers, and edge-cloud hybrid architectures. However, performance assurance for safety-critical AI is still immature. Research explores deterministic execution frameworks and AI-driven workload optimization, but adoption lags outside data center environments.

**Link to Higher Layers**

Execution issues are highly visible: even reliable models (Layer 5) or applications (Layer 7) fail if latency or availability breaks the pipeline. Thus, execution reliability is key to making AI dependable in real-world operations.

### *3.1.9 Layer 9: Monitoring*

The monitoring layer provides observability into AI systems, including logging, anomaly detection, drift detection, and alerting (Aghaei *et al.*, 2025). Its role is to ensure that failures and degradations are identified in time to act.

Typical Monitoring Failures include:

- Silent failure, where monitoring stops without notice.
- Undetected drift, allowing model degradation to persist.
- Alert fatigue, where too many false alarms cause warnings to be ignored.
- Insufficient visibility, leaving blind spots in performance monitoring.

> **Box 10. Cross-Sector Examples of Layer 9: Monitoring**
>
> - **Transportation:** false alerts overwhelm operators, causing them to miss genuine rail system warnings.
> - **Energy:** undetected data drift hides inverter instability.
> - **Healthcare:** clinicians ignore alarms due to over-alerting, missing critical cases.
> - **Manufacturing:** dashboards miss anomalies in machine vibrations due to poor drift detection.
> - **Buildings (HVAC):** monitoring logs silently stop after updates, leaving faults invisible.

Progress includes drift detection algorithms, explainable monitoring dashboards, and federated observability platforms. Yet, industry practice often falls short—many systems lack effective monitoring of AI-specific risks. Emerging research explores self-monitoring AI agents capable of explaining their own uncertainty.

**Link to Higher Layers**

Without reliable monitoring, failures propagate unchecked. This layer serves as the immune system of AI systems, making its reliability critical to long-term safety and trust.

### *3.1.10 Layer 10: Learning*

The learning layer governs adaptation and continuous improvement. It includes retraining pipelines, online learning, reinforcement learning, and feedback integration. Unlike static systems, AI systems evolve after deployment—making this layer uniquely dynamic.

Typical Learning Failures include:
- Concept drift, where learned rules no longer fit current conditions.
- Feedback loop misalignment, where learning amplifies errors.
- Biased retraining data, reinforcing systemic weaknesses.
- Forgetting rare but critical cases during adaptation.

> **Box 11. Cross-Sector Examples of Layer 10: Learning**
>
> - **Transportation:** reinforcement learners mis adapt to seasonal traffic shifts, causing inefficiency.
> - **Energy:** online learning destabilizes grid frequency control.
> - **Healthcare:** auto-updated models degrade after learning from misdiagnoses.
> - **Manufacturing:** retrained classifiers miss rare but costly defects.
> - **Buildings (HVAC):** self-adaptive controllers forget rare emergency conditions, compromising safety.

Research on continual learning, reinforcement learning safety, and robust retraining is advancing rapidly. Industrial practice includes shadow training pipelines and offline validation before deployment (Bayram and Ahmed, 2025), but these safeguards are resource intensive. True safe lifelong learning remains unsolved in critical domains.

**Link to Higher Layers**

Errors in learning undermine trust at the agentic level (Layer 11), as agents adapt in unintended ways. Continuous learning thus transforms reliability from a static property into a moving target that must be actively managed.

### *3.1.11 Layer 11: AI Agent*

The agent layer introduces the highest-level intelligence: reasoning, planning, communication, goal alignment, and multi-agent interaction. This is where AI systems become agentic, acting autonomously to pursue objectives in complex environments.

Typical AI Agent Failures include:
- Goal misalignment, where agent objectives diverge from human intent.
- Multi-agent conflicts, when agents compete instead of cooperating.
- Opaque reasoning, making it difficult to explain or correct decisions.
- Cascading failures, as poor agentic choices amplify lower-layer weaknesses.

> **Box 12. Cross-Sector Examples of Layer 11: AI Agent**
>
> - **Transportation**: autonomous inspection robots' conflict with scheduling agents, causing delays.
> - **Energy**: market bidding agents act competitively, destabilizing grid stability.
> - **Healthcare**: triage agents escalate cases inconsistently, eroding clinician trust.
> - **Manufacturing**: fleets of collaborative robots pursue conflicting optimization goals, creating unsafe conditions.
> - **Buildings (HVAC)**: multiple agents fight over comfort vs. efficiency, leading to unstable building performance.

Agent-based AI is an emerging frontier. Current progress includes goal alignment techniques (e.g., reward shaping, constitutional AI (Sicari *et al.*, 2024)), multi-agent coordination frameworks, and explainable planning methods. However, reliability assurance here is immature, with open challenges around emergent behaviours and human-agent collaboration.

**Link to Higher Layers**

The agent layer sits at the top of the stack, but it reflects vulnerabilities from every lower layer. Poor data (Layer 6) or unstable learning (Layer 10) can distort agent reasoning. Failures here are most visible to society, as they directly affect human trust, safety, and system-level outcomes.

## *3.2 Comparative View: Layered Failure Modes Across AI Paradigms*

The 11-layer failure stack is universal: every AI system — from conventional classifiers to generative models to autonomous agents — operates across these layers. Importantly, failures can occur at any layer. The distinctions drawn here highlight where reliability challenges tend to concentrate or evolve, not where risks exclusively reside (Table 3, Fig. 2).

Thus, the comparative analysis should be understood as showing relative prominence and shifting meanings, rather than the absence of failures in less emphasized layers.

**Conventional AI: Pipeline-Centred Risks**

Conventional AI systems are task-specific and domain-bounded. Reliability risks concentrate in the data–model–application pipeline:
- Layer 5 (Models): overfitting and brittleness under unseen conditions.
- Layer 6 (Data): drift, noise, and labelling errors.
- Layer 7 (Applications): integration failures disrupting workflows.

Other layers remain relevant—for instance, hardware faults (Layer 1) or weak monitoring (Layer 9) — but they are less prominent in defining conventional AI reliability.

**Generative AI: Content-Centred Risks**

Generative AI expands conventional systems by producing open-ended content. This shifts reliability concerns toward accuracy, safety, and appropriateness of generated outputs:
- Layer 5 (Models): hallucinations, controllability, and factual accuracy.
- Layer 6 (Data): reliance on web-scale corpora introduces copyright, bias, and toxicity issues.
- Layer 9 (Monitoring): active filters to detect harmful or misleading content.
- Layer 10 (Learning): risks of misalignment during fine-tuning or feedback-based adaptation.

Lower layers still matter: GPU failures (Layer 1) or orchestration bottlenecks (Layer 8) can undermine generative systems as much as conventional ones.

**Agentic AI: Autonomy-Centred Risks**

Agentic AI combines generative capabilities with planning, reasoning, memory, and multi-agent interaction. Reliability here extends into alignment, safe adaptation, and emergent behaviours:
- Layer 10 (Learning): lifelong adaptation, reinforcement drift, and catastrophic forgetting.
- Layer 11 (AI Agent): goal misalignment, multi-agent conflicts, opaque reasoning, erosion of human trust.

These higher layers intensify the complexity of reliability, but lower layers remain just as critical. A voltage fluctuation (Layer 2) or biased training dataset (Layer 6) can still cascade upward, destabilizing autonomy.

Table 3 Core Risks vs. Paradigm-Specific Emphasis

| Layer | Core Risks (All Paradigms) | Conventional AI Emphasis | Generative AI Emphasis | Agentic AI Emphasis |
|---|---|---|---|---|
| **1 Hardware** | Device wear-out, overheating | Embedded systems | Compute scaling for training/inference | Energy-hungry autonomy at edge |
| **2 Power & Energy** | Voltage instability, depletion | Battery drains in robots | Cloud/edge cost and energy use | Grid-level load from fleets |
| 3 System Software | OS crashes, driver conflicts | Firmware bugs in controllers | Framework dependency fragility | Autonomy requires robust OS for multi-agents |
| 4 Frameworks | Instability, version conflicts | ML pipeline mismatches | Model integration with APIs | Multi-agent orchestration libraries |
| 5 Models | Overfitting, adversarial inputs | Task-specific brittleness | Hallucinations, controllability | Reasoning errors, emergent strategies |
| 6 Data | Drift, noise, mislabelling | Pipeline integrity | Bias in web-scale/synthetic corpora | Self-selected/adaptive data risks |
| 7 Applications | Poor integration, API fragility | Workflow disruption | Copilots, generative design | Autonomous workflows and decision loops |
| 8 Execution | Latency, cold starts, overload | Basic resource scheduling | Inference scaling, real-time streaming | Fleet-level orchestration |
| 9 Monitoring | Silent failure, alert fatigue | Limited logging | Safety filters, toxicity/factuality check | Goal intent and human oversight dashboards |
| 10 Learning | Misaligned retraining, drift | Offline retraining | RLHF misalignment, online fine-tuning | Lifelong adaptation, safe reinforcement |
| 11 AI Agent | Goal conflicts, black-box reasoning | Absent/minimal | Limited prompting autonomy | Full autonomy, multi-agent coordination risks |

Cautionary Note: While certain layers are more visible in one paradigm, all layers remain relevant to reliability. Overlooking foundational or less-prominent layers risks hidden faults cascading upward, where they amplify into system- or society-level consequences. Effective reliability assurance therefore requires vigilance across the entire stack, regardless of paradigm.

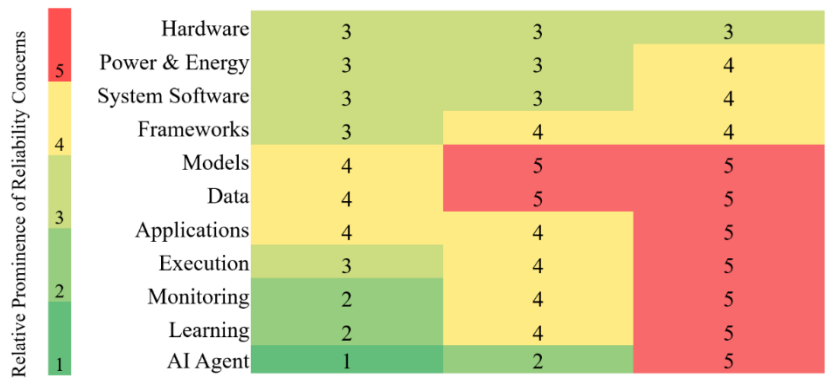

**Fig. 2** Reliability Prominence Across AI Paradigms (11-Layer Stack)

Fig. 2 illustrates how reliability concerns are distributed across the 11-layer failure stack for conventional, generative, and agentic AI systems. Numbers (1–5) represent relative prominence — not the existence of failures, but the degree to which reliability efforts and discussions have typically concentrated at each layer.

In conventional AI, reliability attention has focused most on data, models, and applications (scores 4), reflecting widespread work on data quality, model robustness, and system integration. However, this does not imply that other layers are free of risk; rather, concerns at lower (hardware, power) or higher (monitoring, learning) layers have historically been less emphasized, even though they can produce significant cascading effects.

In generative AI, prominence shifts upward: models and data pipelines reach the highest level (5), as issues like hallucinations, bias, and content safety dominate. Frameworks and monitoring also gain prominence because of the need for decoding stability, safety filters, and real-time oversight.

In agentic AI, emphasis extends further into higher layers. Execution, learning, and agent-level reasoning emerge as top concerns (5), capturing risks such as goal misalignment, unsafe adaptation, and emergent multi-agent conflict.

The progression demonstrates that while all layers remain vulnerable, the center of gravity of reliability attention shifts with paradigm evolution: from mid-stack concerns in conventional AI, to content validity in generative AI, to alignment and coordination challenges in agentic AI.

## 3.3 Paradigm Spotlights: Reliability in Conventional, Generative, and Agentic AI

While all three paradigms share the same 11-layer stack of vulnerabilities, their reliability *personalities* differ in important ways. Section 3.2 highlighted how emphasis shifts across layers. Here, the focus turns to the systemic nature of failures within each paradigm, showing how risks manifest in practice and how the research and standards communities are beginning to respond.

Conventional AI systems, such as predictive classifiers and optimization models, have long been applied in transportation, energy, healthcare, and manufacturing. Their reliability challenges are most visible in the robustness of models, the stability of data pipelines, and the effectiveness of integration with domain-specific applications. Unlike catastrophic failures, breakdowns here often emerge gradually, as performance degrades due to data drift or unanticipated operating conditions. For instance, a railway maintenance system trained on older vibration data may fail to generalize when confronted with new materials, producing false negatives that quietly undermine service availability. To mitigate such risks, the state of the art emphasizes rigorous validation and lifecycle management through ML Ops frameworks, alongside established safety standards such as ISO 26262 for automotive systems (ISO, 2018), IEC 61508 for functional safety (IEC, 2010), and IEEE 1633 for software reliability (IEEE, 2017). Reliability in conventional AI is thus anchored in robustness, validation, and disciplined operational practices.

Generative AI represents a different reliability profile, because it produces open-ended outputs rather than fixed predictions. Here, the central question is not merely whether the model generalizes correctly, but whether its outputs are factually accurate, trustworthy, and safe for downstream use (X. Zhang *et al.*, 2025). This raises risks of hallucination, compounding errors when outputs are recycled into retraining, and misuse of generated content. A healthcare example illustrates the stakes: a generative assistant might produce a convincing but incorrect radiology explanation that persuades clinicians to delay necessary treatment. The problem is not just technical error, but the amplification of error through human trust. To address these risks, current approaches include the NIST AI Risk Management Framework (RMF) (Nist, 2023), which emphasizes transparency, documentation, and safeguards, as well as practical techniques such as red-teaming and output guardrails. The IEEE P7003 standard on algorithmic bias considerations (Koene, Dowthwaite and Seth, 2018) also represents an emerging foundation for governing fairness and reliability in generative models. Reliability in generative AI therefore rests on controlling unpredictability, ensuring factual grounding, and embedding safeguards around open-ended behaviour.

Agentic AI systems extend the reliability challenge even further. By combining reasoning, planning, memory, and interaction across multiple agents, they introduce risks associated with goal alignment, emergent behaviours, and the erosion of human oversight. Failures here are rarely local; instead, they emerge systemically, as autonomous agents optimize local objectives in ways that destabilize global performance. A case in point arises in energy markets, where autonomous bidding agents act rationally in isolation but collectively undermine grid stability. Addressing these risks requires governance as much as technical solutions, including human-in-the-loop oversight mechanisms, ongoing AI alignment research (Russell, 2022), and emerging frameworks such as IEEE P7009 for fail-safe design of autonomous and semi-autonomous systems (IEEE, 2024). In this sense, agentic AI reliability is less about robustness or hallucination and more about managing emergent complexity through alignment, governance, and systemic resilience (Zambare, Thanikella and Liu, 2025).

**Table 4** Comparative Reliability Spotlights Across AI Paradigms

| Paradigm | Systemic Reliability Risks | Illustrative Example | Current Responses / Standards | Key Takeaway |
|---|---|---|---|---|
| **Conventional AI** | Data drift, brittle generalization, integration failures | Railway predictive maintenance misclassifies faults in new materials | ML Ops, ISO 26262, IEC 61508, IEEE 1633 | Reliability anchored in robustness and validation |
| **Generative AI** | Hallucinations, compounding feedback errors, unsafe outputs | Healthcare diagnostic assistant generates misleading radiology explanations | NIST AI RMF, red-teaming, IEEE P7003 | Reliability defined by trustworthiness and safeguards |
| **Agentic AI** | Goal misalignment, emergent behaviours, loss of oversight | Autonomous bidding agents destabilize power grid markets | Human-in-the-loop, AI alignment research, IEEE P7009 | Reliability depends on alignment, governance, and resilience |

Taken together, these spotlights suggest that the trajectory of AI reliability reflects the evolution of paradigms themselves. Conventional AI struggles most with robustness and lifecycle maintenance, generative AI with factuality and safe use of outputs, and agentic AI with alignment and governance of emergent behaviours. Reliability engineering must therefore adapt in step: from strengthening robustness, to embedding safeguards, to governing systemic complexity (Table 4).

## *3.4 Cross-Layer Risks and Cascading Effects*

The 11-layer failure stack shows that vulnerabilities exist at every stage of AI systems, from hardware foundations to agentic coordination. Yet in practice, failures rarely remain confined to a single layer. They propagate across the stack, amplifying risks and creating system-wide consequences. This property distinguishes AI reliability from many traditional engineering contexts, where failures are more localized and predictable (Elder *et al.*, 2024).

Three archetypes of cross-layer propagation are particularly salient:
1. Bottom-up cascades occur when low-level disturbances rise through the stack. For example, noisy sensor data (Layer 6) can degrade model accuracy (Layer 5), which then misguides decision-making applications (Layer 7). In transportation, a mis-calibrated vibration sensor in a railway bogie may trigger false defect predictions, resulting in unnecessary rescheduling and systemic delays.
2. Top-down cascades emerge when high-level decisions or agentic behaviours stress lower layers. In energy systems, bidding strategies by autonomous market agents (Layer 11) may overburden control applications (Layer 7) and even destabilize physical infrastructure (Layer 2). Here, the cascade begins at the reasoning and coordination layer and reverberates downward, producing failures not from material fatigue but from emergent behaviour.
3. Feedback loops are self-reinforcing cycles that blur the boundary between layers. In healthcare, generative models retrained on their own outputs (Layer 10) can inject bias into data pipelines (Layer 6), which in turn worsens model reliability (Layer 5). What begins as a small drift compound into systemic diagnostic inaccuracies that erode clinician trust (Layer 11).

These patterns illustrate why AI reliability cannot be reduced to guarding individual components. Cascading failures amplify risks and often manifest in surprising ways — turning local disturbances into global instability. They also highlight a gap in current organizational practice: many stakeholders can identify obvious model or application failures, but fewer recognize the cross-layer dynamics that cause small faults to grow into systemic risks.

This gap motivates the second framework of the chapter: awareness mapping. By evaluating how organizations perceive and respond to layered and cross-layer risks, awareness mapping provides a maturity-oriented lens for designing dependable AI architectures.

## 4. Awareness Mapping: From Failure Modes to Reliability Maturity

Reliability in AI systems is not only a matter of technical safeguards; it also depends on how well individuals and organizations perceive, understand, and prepare for risks. Vulnerabilities span all 11 layers of the failure stack, yet their impact is strongly shaped by awareness. In practice, awareness is uneven: most practitioners can recognize failures at the model or application layer, but far fewer anticipate risks in data pipelines, adaptive learning dynamics, or multi-agent interactions—where some of the most consequential failures arise.

To address this gap, we introduce awareness mapping: a structured approach for assessing how comprehensively organizations understand reliability risks across AI systems. Awareness mapping shifts the focus from failures themselves to the recognition of those failures, positioning awareness as a critical dimension of reliability. More importantly, it provides an evidence-based foundation for strategic decision-making, helping organizations review their current preparedness, identify blind spots, and prioritize improvements in governance, training, monitoring, and lifecycle management. In this way, awareness mapping serves both as a diagnostic lens and as a practical tool for shaping reliability strategies in conventional, generative, and agentic AI.

This section develops the framework in four steps. Section 4.1 explains how failures across the 11-layer stack are operationalized into specific reliability issues that can be scored as points of awareness. Section 4.2 introduces a five-level maturity scale that translates these scores into stages of organizational readiness. Section 4.3 presents empirical insights from practitioner surveys, highlighting current blind spots and uneven awareness across sectors. Finally, Section 4.4 links awareness mapping to Dependability-Centred Asset Management (DCAM), showing how awareness maturity can guide lifecycle strategies for trustworthy and resilient AI.

Ultimately, awareness mapping transforms reliability from a reactive concern into a proactive strategic capability, enabling organizations to anticipate, govern, and continually improve AI systems in step with technological evolution.

### *4.1 Scoring Awareness Across Failure Modes*

Section 3 catalogued the technical vulnerabilities of AI systems through an 11-layer failure stack. To move from technical vulnerabilities to measurable organizational maturity, we link each failure mode to its corresponding body of reliability studies. The idea is that awareness is not simply knowing that failures may occur, but being familiar with methods, studies, or practices that address them. Each study area therefore becomes a potential *awareness point*.

In our implementation, the 11-layer stack is mapped onto approximately 47 reliability studies drawn from conventional engineering, AI safety research, and emerging generative/agentic AI work. For example:
- Hardware failures such as memory corruption are linked to studies on ECC diagnostics and thermal aging analysis.
- Data-related risks are linked to methods for drift detection, data pipeline validation, and bias auditing.
- Model-level concerns correspond to adversarial robustness testing, uncertainty quantification, or hallucination suppression.
- At the agentic level, studies focus on alignment verification, simulation-in-the-loop evaluation, and multi-agent stress testing.

Table 5 summarizes this mapping. For each layer of the stack, it distinguishes between:
- Baseline reliability studies in conventional AI systems,
- Additional focus areas introduced by generative AI, and
- New reliability challenges and studies specific to agentic AI.

When used in practice, respondents are asked to indicate which of these study areas they are aware of. Each affirmative response counts as one point toward an overall awareness score. The total score thus reflects not just recognition of risks, but also familiarity with concrete methods for addressing them.

This scoring approach — assigning one point per identified study area, across 47 in total — is deliberately simplified. It does not capture the depth of knowledge or implementation quality. However, it provides a transparent and actionable baseline: organizations can benchmark their maturity, identify blind spots, and prioritize capacity building.

Section 4.2 builds on this by introducing a five-level maturity scale that translates awareness scores into stages of organizational readiness.

**Table 5** Reliability Studies Across Layers of Conventional, Generative, and Agentic AI Systems

| Layer | Conventional AI – Reliability Studies | Generative AI – Added Reliability Studies | Agentic AI – Added Reliability Studies |
|---|---|---|---|
| **1. Hardware** | 1. GPU/TPU diagnostics; 2. ECC error analysis; 3. interconnect reliability; 4. thermal aging studies; 5. vibration stress testing | Long-run accelerator reliability; mixed-precision error validation; sustained training/inference stress tests | Robotics hardware robustness; actuator/sensor degradation analysis; fleet-level redundancy evaluation |
| **2. Power & Energy** | 6. Voltage fluctuation testing; 7. surge/UPS resilience; 8. thermal envelope validation; 9. blackout/brownout recovery | Energy efficiency studies for large-scale training; power-capping effects on QoS; cooling/thermal resilience in datacenters | Battery SoH/SOC forecasting; energy-aware autonomy validation; safe degradation pathways under power loss |
| **3. System Software** | 10. Kernel panic forensics; 11. container/VM stability tests; 12. firmware update safety; 13. clock skew validation | CUDA/ROCm compatibility matrices; NUMA/I/O contention studies; GPU memory allocation fault analysis | RTOS determinism studies; watchdog & safety monitor reliability; secure update protocols in edge/IoT settings |
| **4. AI Frameworks** | 14. Dependency conflict resolution; 15. build reproducibility testing; 16. deterministic training benchmarking | Tokenizer stability; decoding reproducibility; safety filter integration validation | Agent-framework reliability (tool use contracts, sandboxing); plug-in orchestration correctness studies |
| **5. Models** | 17. OOD robustness testing; 18. calibration metrics; 19. adversarial robustness studies; 20. concept drift monitoring; 21. uncertainty quantification | Hallucination detection benchmarks; factuality/toxicity red-teaming; controllability experiments (prompt constraints, decoding strategies) | Reasoning & planning fidelity testing; goal alignment verification; emergent strategy audits in multi-agent setups |
| **6. Data** | 22. Label quality audits; 23. data drift & leakage detection; 24. data pipeline validation; 25. synthetic data robustness; 26. provenance/lineage verification | Web-scale bias detection; PII/copyright filtering studies; synthetic data robustness analysis | Memory/log data integrity; self-generated interaction data validation; agent feedback loop consistency checks |
| **7. Applications** | 27. API failure recovery analysis; 28. latency/throughput impact studies; 29. user trust calibration 30. HMI resilience evaluation | Guardrail UI testing; human trust calibration studies; post-processing verification frameworks | Governance of end-to-end workflows; safe-abort/approval mechanisms; role/permission modelling validation |
| **8. Execution** | 31. Chaos engineering; 32. Deployment fault injectioion; 33. orchestration stress tests; 34. autoscaling & rollback analysis; 35. environment drift control | KV-cache stability; batching/streaming trade-off benchmarking; heterogeneous accelerator scheduling | Real-time multi-agent orchestration testing; consensus/coordination stress tests; runaway loop prevention |
| **9. Monitoring** | 36. Observability coverage audits; 37. anomaly detection benchmarks; 38. drift detection latency analysis; 39. alert fatigue mitigation | Hallucination/toxicity monitoring frameworks; content safety dashboards; multi-pass critique/verification pipelines | Goal deviation monitors; plan-conformance validation; safety sentinels for multi-agent monitoring; escalation pathways |
| **10. Learning** | 40. Online learning romustness 41. Offline retraining hygiene studies; 42. drift-adaptive training validation; 43. rollback of regressions | RLHF/RLAIF stability; online fine-tuning safety validation; preference shift monitoring | Lifelong learning safety validation; safe exploration benchmarks; catastrophic forgetting protection; policy versioning safeguards |
| **11. AI Agent** | — (minimal in conventional AI) | Prompt-loop guardrail validation; limited autonomy safety checks | Autonomy safety cases; alignment verification frameworks; simulation-in-the-loop testing; multi-agent game-theoretic stress tests |
| **Cross-cutting** | 44. Silent failure detection studies; 45. cascading-fault modelling; 46. multi-layer resilience co-design; 47. security breach defense | Red-teaming suites for generative models; usage policy enforcement; provenance/auditability of generated outputs | Governance & oversight frameworks; auditability of plans/actions; inter-agent safety norms; societal/market impact constraints |

Table 5 translates the 11-layer failure stack into a set of reliability studies that serve as awareness points. For each layer, the baseline column reflects established reliability practices in conventional AI, while the generative and agentic columns extend these with additional concerns unique to open-ended content generation and autonomous reasoning. Each study corresponds to a point of awareness: if a practitioner or organization can recognize and articulate the relevance of that study to their system, it contributes one point to their awareness score. In total, the 47 identified studies (baseline + extensions) define the assessment space.

These awareness points form the backbone of the scoring method described in Section 4.2, where the total number of recognized points (0–47) is mapped onto a five-level maturity scale. This mapping transforms awareness from a qualitative impression into a structured diagnostic tool, enabling organizations to benchmark their preparedness, identify blind spots, and prioritize improvements in AI reliability strategies.

## 4.2 Awareness Maturity Levels

The awareness score provides a structured way to measure how reliably individuals or organizations perceive risks across the AI system stack. It is derived from the 47 diagnostic points identified in Section 4.1, corresponding to failure modes and reliability studies spanning all 11 layers—from hardware and data pipelines to adaptive learning and agentic reasoning. Each point represents a distinct reliability concern. Respondents are asked whether they are aware of each issue; every positive response contributes one point, producing a maximum possible score of 47. The total reflects the breadth of reliability risks that stakeholders consciously recognize.

It is important to emphasize that these **47 points are drawn from conventional AI systems**. They capture well-established risks such as hardware degradation, power instability, data drift, model overfitting, or monitoring blind spots. However, the framework is designed to be **extensible**. By integrating generative-specific risks (e.g., hallucinations, factuality errors, unsafe content) and agentic-specific risks (e.g., goal misalignment, reasoning drift, emergent multi-agent conflicts), the same approach can evolve into a more comprehensive awareness instrument. In this sense, the 47 points should be seen as a **baseline example**, not a closed set.

To make scores meaningful, we map them onto a five-level maturity scale (Table 6). This scale captures not only the number of risks recognized, but also the organizational posture implied by that recognition:

- Level I (0–9 points): Unaware

Organizations at this stage have not considered AI reliability in a systematic way. Only obvious failures such as application crashes or output errors are noticed. Reliability is absent from conversations and practices, leaving systems highly exposed.

- Level II (10–19 points): Fragmented Awareness

Some failures are acknowledged, typically at the model or application level, but without systematic measures. Risks in deeper layers—such as data pipelines, execution, or monitoring—remain overlooked. Reliability efforts are piecemeal and reactive.

- Level III (20–29 points): Emerging Multi-Layer Awareness

Awareness expands across multiple layers. Teams recognize that AI can fail beyond surface-level errors and begin to consider mitigation strategies. Data quality, model robustness, and infrastructure reliability enter the discussion, though responses remain mostly reactive and uneven.

- Level IV (30–39 points): Proactive Systemic Awareness

Reliability is actively monitored and tested across several layers. Governance mechanisms begin to take shape, supported by structured dependability practices. Blind spots remain—particularly in higher-order risks such as goal alignment or emergent multi-agent behaviors—but awareness is no longer confined to isolated issues.

- Level V (40–47 points): Comprehensive Cross-Layer Reliability

**Table 6** summarizes these levels, their score ranges, and their practical meaning.

| Level | Score Range | Descriptor | Meaning in Practice |
|---|---|---|---|
| **I. Unaware** | 0 – 9 | No consideration | Reliability absents from discussions; only obvious failures (e.g., crashes) noticed. |
| **II. Fragmented Awareness** | 10 – 19 | Isolated recognition | Failures acknowledged mainly at model or application level; no systematic measures for deeper risks. |
| **III. Emerging Multi-Layer Awareness** | 20 – 29 | Expanding recognition | Failures at several layers are acknowledged; some mitigations applied, though mostly reactive. |
| **IV. Proactive Systemic Awareness** | 30 – 39 | Structured approach | Reliability monitored and tested across multiple layers; governance mechanisms beginning to emerge. |
| **V. Comprehensive Cross-Layer Reliability** | 40 – 47 | Full integration | Reliability strategy spans all 11 layers, including generative and agentic risks; embedded in organizational culture. |

Awareness spans virtually all diagnostic points, including those in reasoning, adaptation, and agentic coordination. Reliability is embedded in organizational culture, supported by systematic dependability engineering, proactive governance, and lifecycle strategies.

This scoring method is deliberately simplified. It privileges breadth of recognition (how many failure types are known) over depth of understanding (how well those failures are mitigated). As such, it should be interpreted as an indicative measure of organizational maturity rather than a definitive evaluation of capability. Its strength lies in comparability: by using the same scoring approach across teams, sectors, or domains, it highlights blind spots, benchmarks progress, and guides interventions.

Ultimately, this maturity framework turns awareness from a static description into a developmental roadmap. Progression from Level I to Level V reflects not only wider recognition of risks, but also deeper institutional capacity to anticipate, mitigate, and continually improve reliability across conventional, generative, and agentic AI systems.

### *4.3 Empirical Insights from Practice*

To assess how the awareness mapping framework performs in practice, we applied it during several invited keynote talks across domains including transportation, energy, and manufacturing. In these sessions, participants — primarily engineers, managers, and researchers engaged with AI-enabled systems — were asked to complete a rapid diagnostic exercise. They were presented with a list of 47 reliability issues mapped to the 11-layer failure stack (see Section 3) and asked to indicate which ones they had previously considered in their work. Each identified issue counted as one point toward an overall awareness score.

The results were strikingly consistent. While participants readily identified a handful of visible issues — such as model overfitting, data noise, or application downtime — far fewer recognized vulnerabilities at deeper layers such as power stability, execution latency, or adaptive learning drift. Scores clustered toward the lower end of the maturity ladder, with most individuals and organizations falling within Level I (Unaware) or Level II (Fragmented Awareness). Only a minority reached Level III, and very few respondents demonstrated the cross-layer recognition required for Levels IV or V.

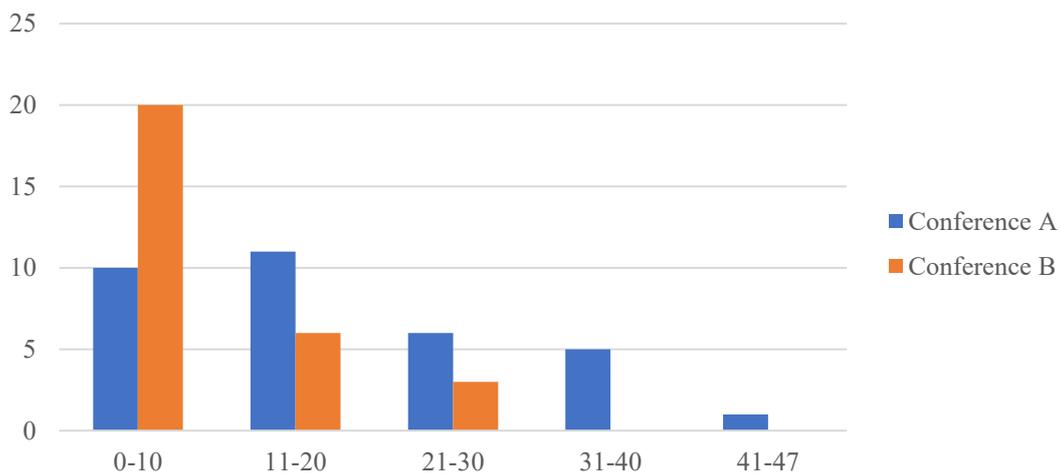

**Fig. 3** Distribution of awareness scores from keynote session on AI reliability in the physical asset management and transportation sector. Conference A: 30th International and Mediterranean HDO (MeditMaint2025); HDO: Croatian Maintenance Society) conference, from 19th -22nd May, Rovinj, Croatian. Conference B: 1st International Conference on Transportation Systems (TS2025), from 16th -18th June 2025, Lisbon, Portugal.

Fig. 3 illustrate the distribution of scores from two keynote sessions. Despite differences in sectoral focus, both show a similar skew: most participants recognized fewer than 20 of the 47 failure modes, with only a very small group achieving awareness scores above 30. Notably, both surveys confirmed the same misalignment: failures are most often noticed at the model and application layers, while the most consequential risks — such as cascading effects from data pipelines, execution environments, or agentic coordination — are rarely acknowledged.

This evidence underscores two lessons. First, organizations may be systematically underestimating reliability risks by concentrating on surface-level failures. Second, even a simple diagnostic instrument such as the 47-item checklist can serve a dual role: not only measuring awareness but also expanding it. Participants often reported that encountering failure modes outside their prior experience — particularly in domains such as power and energy management or multi-agent coordination — reshaped their perspective on where reliability strategies should focus.

Taken together, these findings highlight both the urgency of broadening reliability awareness and the potential of structured mapping to support this expansion in practice.

### *4.4 Linking Awareness to Dependability-Centred Asset Management (DCAM)*

The awareness mapping framework is more than a diagnostic tool for measuring how organizations perceive AI reliability risks. It also aligns directly with the broader paradigm of Dependability-Centred Asset Management (DCAM), which extends traditional asset management by embedding reliability, availability, maintainability, and safety (RAMS) across the lifecycle of both physical and digital assets (Lin, 2025). In this view, AI systems — whether conventional, generative, or agentic — are treated as critical organizational assets whose dependability must be managed alongside physical infrastructure.

Awareness levels serve as an entry point into DCAM practice. Organizations at Level I or II lack the readiness to integrate AI reliability into asset management strategies, as risks remain unrecognized and decision-making is reactive. By Level III, awareness expands across multiple layers, allowing AI reliability considerations to begin influencing operational strategies such as monitoring, anomaly detection, and lifecycle planning.

At Level IV, systemic awareness begins to merge with DCAM structures. AI reliability is embedded into governance and management processes, including cross-layer monitoring, predictive maintenance supported by AI diagnostics, and structured validation protocols. At the highest level, Level V, awareness is comprehensive and coupled with institutionalized dependability practices. Here, DCAM and awareness mapping converge: reliability strategies become proactive, embedded into design, deployment, and continuous improvement, spanning both physical assets and AI-driven systems.

Crucially, awareness mapping provides a diagnostic bridge between technical AI reliability and organizational maturity in asset management. It reveals blind spots that managers may overlook, guiding targeted interventions. For example, a utility operator who recognizes model-level errors, but neglects execution-layer risks may fail to secure grid stability under dynamic demand; awareness mapping directs attention to these gaps and extends DCAM practices into digital infrastructures.

In this way, awareness mapping enriches DCAM in two complementary ways:
1. By embedding AI reliability explicitly within the asset management lifecycle.
2. By offering a structured roadmap for organizations to progress from fragmented awareness to comprehensive dependability.

Together, the two frameworks extend the scope of asset management to meet the challenges of increasingly autonomous, adaptive, and agentic AI systems, ensuring that reliability is not an afterthought but a core principle of lifecycle strategy.

### 5. Conclusion and Outlook

This chapter has advanced a dual contribution to the study of AI reliability. First, it introduced the 11-layer failure stack — a structured framework for tracing vulnerabilities across conventional, generative, and agentic AI systems. By moving from physical computation layers to reasoning and agentic coordination, the stack highlights that reliability challenges are not confined to models or applications but extend throughout the entire lifecycle of AI systems. Case vignettes illustrated how failures at seemingly minor layers can cascade into systemic disruptions, underscoring the need for a cross-layer perspective.

Second, the chapter developed the concept of awareness mapping — a method for evaluating how well organizations recognize and prepare for reliability risks. By translating failure modes into measurable awareness points, the framework provides a diagnostic lens into organizational maturity. The five-level maturity scale helps practitioners identify blind spots, benchmark progress, and guide the integration of reliability considerations into strategy and governance. The alignment of awareness mapping with Dependability-Centred Asset Management (DCAM) further shows how reliability can be embedded within broader lifecycle practices for both physical and digital assets. Together, the failure stack and awareness mapping demonstrate that reliability is not a fixed property but a moving target, shaped by technological evolution and organizational readiness. Conventional AI, generative AI, and agentic AI each bring distinct vulnerabilities, yet the principle is consistent: safeguarding reliability requires both technical safeguards and organizational awareness, reinforced across layers and over time.

Looking forward, several avenues emerge. Future work could refine awareness mapping beyond breadth of recognition to capture depth of understanding and mitigation capacity. Comparative studies across industries would provide empirical evidence on sector-specific blind spots and resilience strategies. In parallel, the failure stack could evolve as new paradigms — such as embodied intelligence or hybrid human–AI collectives — introduce additional layers of complexity. Finally, integration with standards, regulation, and governance frameworks will be essential to ensure that awareness and dependability principles scale with the societal deployment of generative and agentic AI.

In conclusion, layered failure analysis and awareness mapping together offer a foundation for moving from reactive responses to proactive reliability strategies. By situating AI reliability within a systemic, lifecycle-oriented framework, they provide both a diagnostic lens and a roadmap — supporting the trustworthy, sustainable, and safe deployment of AI in increasingly critical domains.